\documentclass[a4paper,11pt]{article}
\usepackage{pos}
\usepackage{graphicx}  % needed for figures
\usepackage{dcolumn}   % needed for some tables
\usepackage{bm}        % for math
\usepackage{standalone}
\usepackage{enumitem}
\usepackage[absolute,overlay]{textpos}
\usepackage{xcolor}
\usepackage{slashed}
\usepackage{booktabs}
\usepackage{multirow}
\usepackage{amsmath}
\usepackage{bbm}
\usepackage{bbold}
\usepackage{stackrel}
\usepackage{rotating}
\usepackage{pifont}
\usepackage{mathtools}
\usepackage{pgfplots}
\usepackage{tikz}
\usepackage{nicefrac}
\usetikzlibrary{calc}

\definecolor{myred}{RGB}{227,0,31}
\definecolor{myblue}{RGB}{56,125,216}
\definecolor{mygreen}{RGB}{84,122,58}
\usepackage{hyperref}
\hypersetup{
    colorlinks=true,       % false: boxed links; true: colored links
    linkcolor=blue,          % color of internal links
    citecolor=blue,        % color of links to bibliography
    filecolor=blue,      % color of file links
    urlcolor=blue           % color of external links
}
\usepackage{simplewick}

\usepackage{float} 
\usepackage{tikz}
\usepackage{xspace}

\usepackage{subfig}

\title{Smoothing properties of the Wilson flow and the topological charge}
\ShortTitle{Smoothing properties of the Wilson flow and the topological charge}

\author[a,b]{Michele~Della~Morte}
\author*[a,b,c]{Benjamin~J{\"a}ger}
\author[a,b]{Sofie~Martins}
\author[d]{J.~Tobias~Tsang}

\affiliation[a]{Dept. of Mathematics and Computer Science, University of Southern Denmark, Campusvej 55, 5230 Odense M, Denmark}
\affiliation[b]{Quantum Theory Center ($\hbar$QTC), University of Southern Denmark, Campusvej 55, 5230 Odense M, Denmark}
\affiliation[c]{Danish Institute for Advanced Study, University of Southern Denmark, Campusvej 55, 5230 Odense M, Denmark}
\affiliation[d]{Department of Theoretical Physics, CERN, 1211 Geneva 23, Switzerland}

\emailAdd{jaeger@imada.sdu.dk}

\abstract{

We study SU$(N_C)$ gauge theories with a single fermion in the two-index antisymmetric representation to predict the mesonic spectrum of supersymmetric $\mathcal{N}=1$ SYM theories. Using gradient flow methods, we investigate fractional topological charges in $N_C = 4$ ensembles with varying lattice spacings. We show that the use of overimproved gauge actions (specifically the DBW2 action) in the
smearing kernel stabilises the values of the topological charge already at moderate values of the flow time, while this is not the case for the standard Wilson flow.

\begin{textblock}{20}(15.0,1.70)
CERN-TH-2025-021\\
\end{textblock}% 
}

\FullConference{The 41st International Symposium on Lattice Field Theory (LATTICE2024)\\
 28 July - 3 August 2024\\
Liverpool, UK\\}

\begin{document}
\maketitle

\section{Introduction}

We explore the large-$N_C$ limit of a one-flavour theory with a single fermion in the two-index antisymmetric representation as a proxy to predict the low-lying mesonic spectrum of supersymmetric $\mathcal{N}=1$ SYM theories~\cite{Ziegler:2021nbl,Jaeger:2022ypq,DellaMorte:2023ylq,Martins:2023kcj}. This study builds on the work of Corrigan and Ramond (CR)~\cite{Corrigan:1979xf}. In such theories, gauge configurations with non-integer winding numbers may arise. However, near the continuum limit and using periodic boundary conditions (PBC), it has been shown that a geometrical quantity with the properties of a topological charge can be constructed, which only takes integer values~\cite{Luscher:1981zq}. In contrast, fractional charges appear when 't Hooft twisted boundary conditions are imposed~\cite{THOOFT1979141,GarciaPerez:2000aiw,Gonz_lez_Arroyo_2020}. 

In~\cite{Fodor:2009nh}, the ``disappearance'' of fractional charges toward the continuum limit with PBC was investigated numerically for the sextet theory (SU(2) gauge group with fermions in the two-index symmetric representation) in the quenched approximation. We aim to repeat such a study for $SU(N_C)$ gauge theories with one dynamical quark in the two-index antisymmetric representation. Differently from the study in~\cite{Fodor:2009nh}, which relied on the Atiyah-Singer index theorem~\cite{Atiyah:1968mp}, we define the topological charge by smoothing out the gauge configurations using the gradient flow, as proposed in~\cite{Luscher:2010iy}. To this end, we generated four ensembles (see Table~\ref{tab.over}) for $N_C=4$ with different lattice spacings but otherwise (approximately) fixed physical parameters to investigate whether fractional charges are present in such theories.

At first glance, this task may seem straightforward: apply the gradient flow to the gauge configurations and verify whether fractional topological charges appear. However, we found that defining the topological charge in this context is more challenging than initially expected.

In the following, we compare the ``standard'' Wilson flow~\cite{Luscher:2010iy} and the DBW2 flow (doubly blocked from the Wilson action in two coupling space)~\cite{QCD-TARO:1999mox}. In general, the flow equation is defined as
\begin{equation}
    \frac{d}{d t} U_{x,\mu}(t) = - \frac{\delta}{\delta U_{x,\mu}} S_\mathrm{flow}(U_{x,\mu}(t)),
\end{equation}
where $\frac{\delta}{\delta U_{x,\mu}}$ is the Lie derivative with respect to the gauge field. The flow kernel $S_\mathrm{flow}$ is given by the sum of plaquettes and rectangles with appropriate coefficients:
\begin{equation}
    S_\mathrm{flow} = c_0 S_\mathrm{plaquette} + c_1 S_\mathrm{rectangle}.
\end{equation}
The standard Wilson flow corresponds to $c_0 = 1$, $c_1 = 0$, while the DBW2 flow uses $c_0 = 12.2704$, $c_1 = -1.4088$~\cite{QCD-TARO:1999mox}. For the topological charge, we use the clover definition~\cite{Alexandrou:2017hqw}:
\begin{equation}
    Q = \sum_x q(x) = \frac{1}{16 \pi^2} \sum_x \epsilon_{\mu \nu \rho \sigma} \mathrm{Tr} \left( C^\mathrm{Clover}_{\mu \nu}(x) C^\mathrm{Clover}_{\rho \sigma}(x) \right).
\end{equation}
To measure the smoothness of our flowed gauge configurations
we use the definition from~\cite{Luscher:2010iy}
\begin{equation}
    h = \underset{p}{\mathrm{max}}\, \mathrm{Re}\, \mathrm{Tr} \left( \mathbb{1} - V_t (p)\right). 
    \label{eq:h}
\end{equation}
where $V_t(p)$ is given by the product of the smeared link variables around $p$.
The maximum is taken over all plaquettes $p$. The (space-time) average can be defined analogously.  

\begin{figure}%[!h]
\centering
\subfloat{
\includegraphics[width=0.499\textwidth]{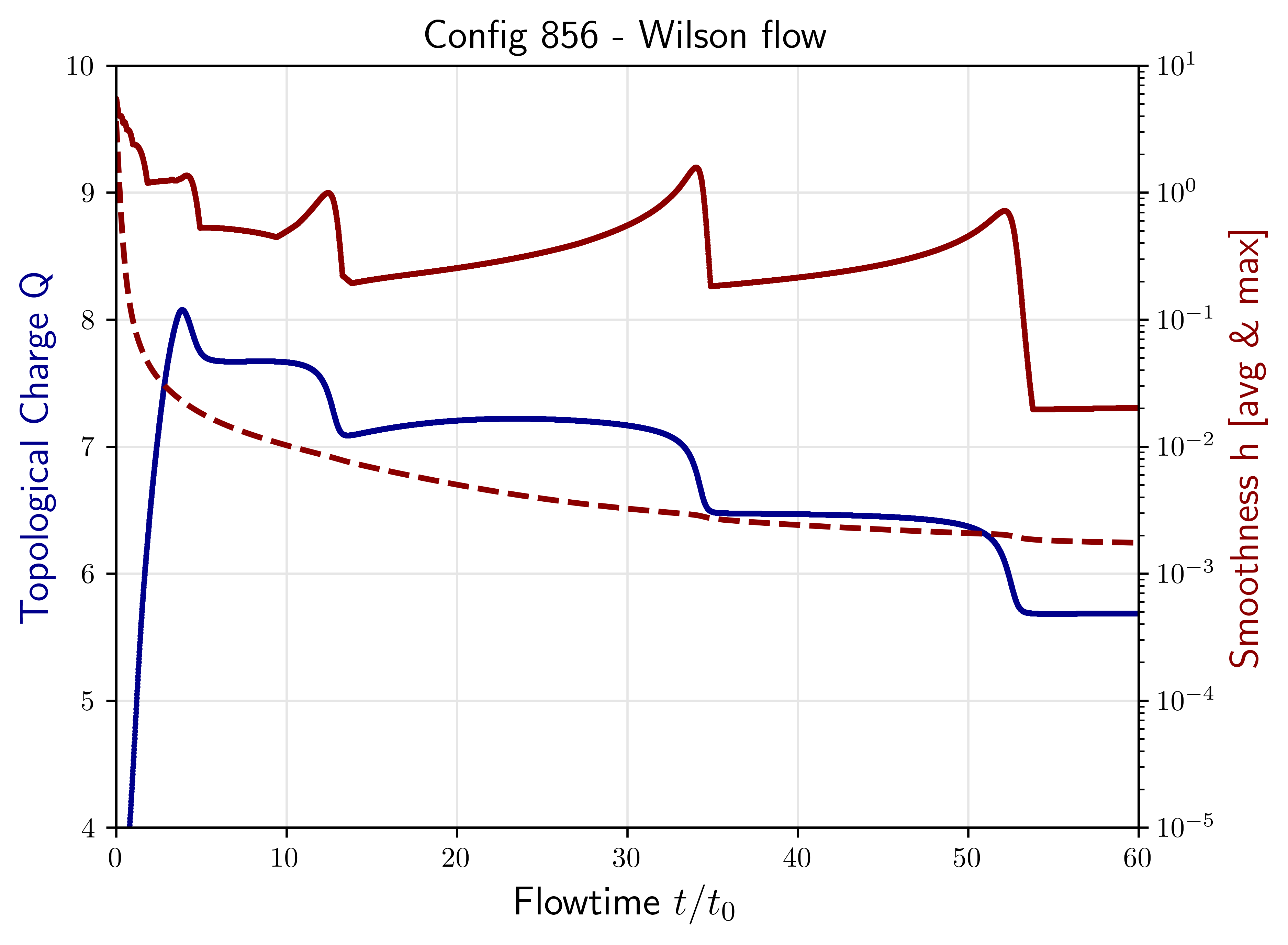}
}
\subfloat{
\includegraphics[width=0.499\textwidth]{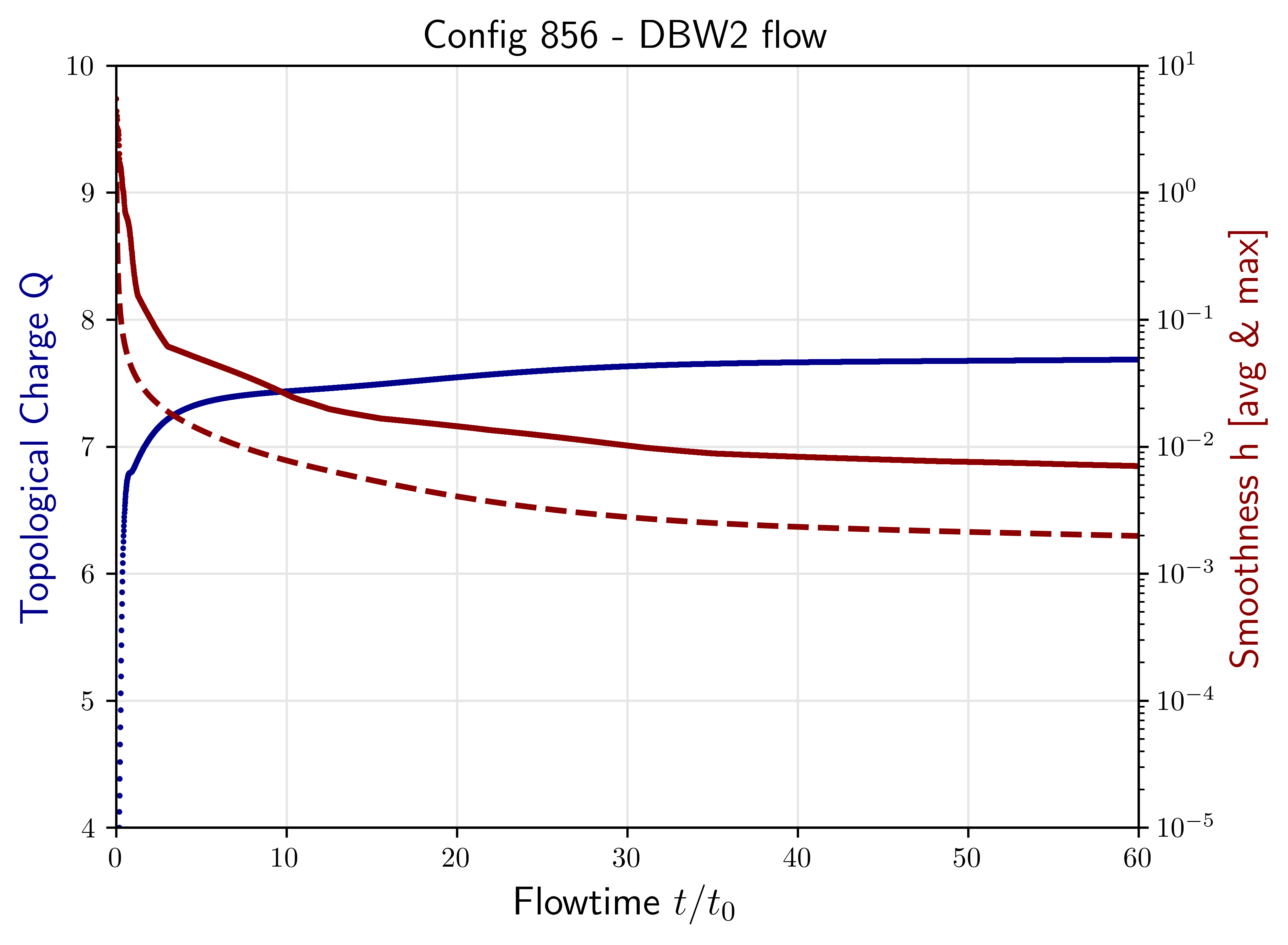}
}

\caption{Topological charge $Q$ (in blue) as a function of the flow time using the Wilson flow (left) and DBW2 flow (right) and the average (in red, dashed) and maximum (in red) of the smoothness parameter $h$ (see eq.~\ref{eq:h} and below) for both.
Results from the $\beta=7.1$ ensemble in Table~\ref{tab.over}.}
\label{fig.example}
\end{figure}

The left panel of figure~\ref{fig.example} illustrates results from the gradient flow using the plaquette for the flow evolution (commonly referred to as the Wilson flow).
The right panel shows results for the DBW2-smearing. In the first case the value of the topological charge keeps jumping also for rather large values of $t$, making it difficult to define a consistent value.  The jumps coincide with large values of the parameter $h$, which is not a monotonic function of $t$. The average value of $\mathrm{Re}\, \mathrm{Tr} \left( \mathbb{1} - V_t (p)\right)$ instead decreases monotonically with $t$, making it clear that the Wilson flow provides in fact, {\it on average}, a smoothing procedure.
In view of that it is conceivable that the jumps in $h$ are caused by large fluctuations at the lattice scale (e.g., instantons of very small size).
This issue has been known for some time in the similar context of cooling and (over)improved kernels have been proposed to address it~\cite{GarciaPerez:1993lic}. More recently, the stability of topological sectors has been studied for pure gauge SU(2) in~\cite{Tanizaki:2024zsu}. The right panel of figure~\ref{fig.example} supports the 
conclusions of such studies.
\section{Lattice Setup}

For our numerical study, we use a single flavour of two-index antisymmetric fermions with the L\"uscher-Weiz gauge action. We adopt the Wilson formulation with tree-level improvement for the fermions ($c_{sw} = 1$). We generate four ensembles at different $\beta$ values, ranging from $7.1$ to $7.4$. The hopping parameter $\kappa$ is chosen such that the flavour-diagonal connected pseudo-scalar mass (CPS) is approximately constant across different gauge couplings. An overview of the ensembles is provided in Table~\ref{tab.over}.
\begin{table}%[!h]
    \centering
    \begin{tabular}{c|ccccc}
    \toprule
$\beta$ & 7.1 & 7.2 & 7.3 & 7.4  \\
$\kappa$  & 0.15770 & 0.15651 & 0.15525 & 0.15401 \\
$L/a$      & 12   & 14   & 16   & 18    \\
$N_{\mathrm{cnfg}}$ & 648 & 346 & 640 & 388\\
    \toprule
$a$ [fm] & $0.171(1)$ & $0.136(1)$ & $0.117(1)$ & $0.103(1)$\\
$L$ [fm] & $2.052(2)$ & $1.904(2)$ & $1.872(2)$ & $1.854(2)$\\
$m_\mathrm{CPS}$ [MeV] & $748(2)$ & $740(3)$ & $734(3)$ & $756(4)$ \\
$m_\mathrm{CPS} L$& $7.78(5)$ & $7.14(6)$ & $6.96(7)$ & $7.10(8)$ \\
$t_0/a^2$ & $0.87(1)$ & $1.37(1)$ & $1.86(1)$ & $2.39(1)$\\
\bottomrule 
    \end{tabular}
    \caption{Overview of parameters of the lattice ensembles. Physical volumes and connected pseudo-scalar masses $(m_\mathrm{CPS})$ are matched to ensure an approximate  line of constant physics. The errors are purely statistical.}
    \label{tab.over}
\end{table}
This setup is unphysical, with physical values included only for comparison with other lattice simulations and to gauge the scales involved. To set the scale, we use the condition
\begin{equation}
    \langle E_\mathrm{plaq} \rangle \, t_0^2 = 0.422,
\label{eq:t0}
\end{equation}
where $E_\mathrm{plaq}$ is the plaquette definition of the energy density introduced in~\cite{Luscher:2010iy}.
The equation above is obtained by rescaling the standard definition of $t_0$ adopted in QCD (namely $\langle E \rangle t_0^2=0.3$) using the leading dependence of $\langle E \rangle$ on $N_C$ for fixed 't Hooft coupling. That results in a $(N_C^2-1)/N_C$ scaling for the r.h.s. of eq.~\ref{eq:t0}.
For comparison with other simulations, we use
\begin{equation}
    \sqrt{8 t_0^\mathrm{ref}} = 0.45 \, \mathrm{fm},
\end{equation}
which is an average of the $N_f = 0$~\cite{Luscher:2010iy} and $N_f = 2$~\cite{Bruno:2013gha} results in QCD.

The simulations were performed using the HiRep code~\cite{PhysRevD.81.094503}, recently optimised for GPUs~\cite{Martins:20248+,Martins:2024sdd} and we have been using the git-hash ca4aa637. The code is publicly available, and input files are provided in the arXiv submission. To reduce autocorrelation, gauge configurations were separated by eight MDUs with a trajectory length of $\tau = 2.0$. The integration steps of the hybrid Monte Carlo algorithm were tuned for an acceptance rate exceeding 90\%, and every fourth configuration was stored. To ensure thermalisation, the first 800 MDUs were discarded. For the flow measurements we use a third level integrator~\cite{Luscher:2010iy} with a stepsize of $\epsilon = 0.01$ for a maximum flowtime of 64 in lattice units. All configurations were generated on the GPU partition of the LUMI supercomputer, whereas the flow measurements were performed on  CPU partitions (LUMI and Discoverer).

\section{Results}

\begin{figure}%[!ht]
\centering
\includegraphics[width=1.0\textwidth]{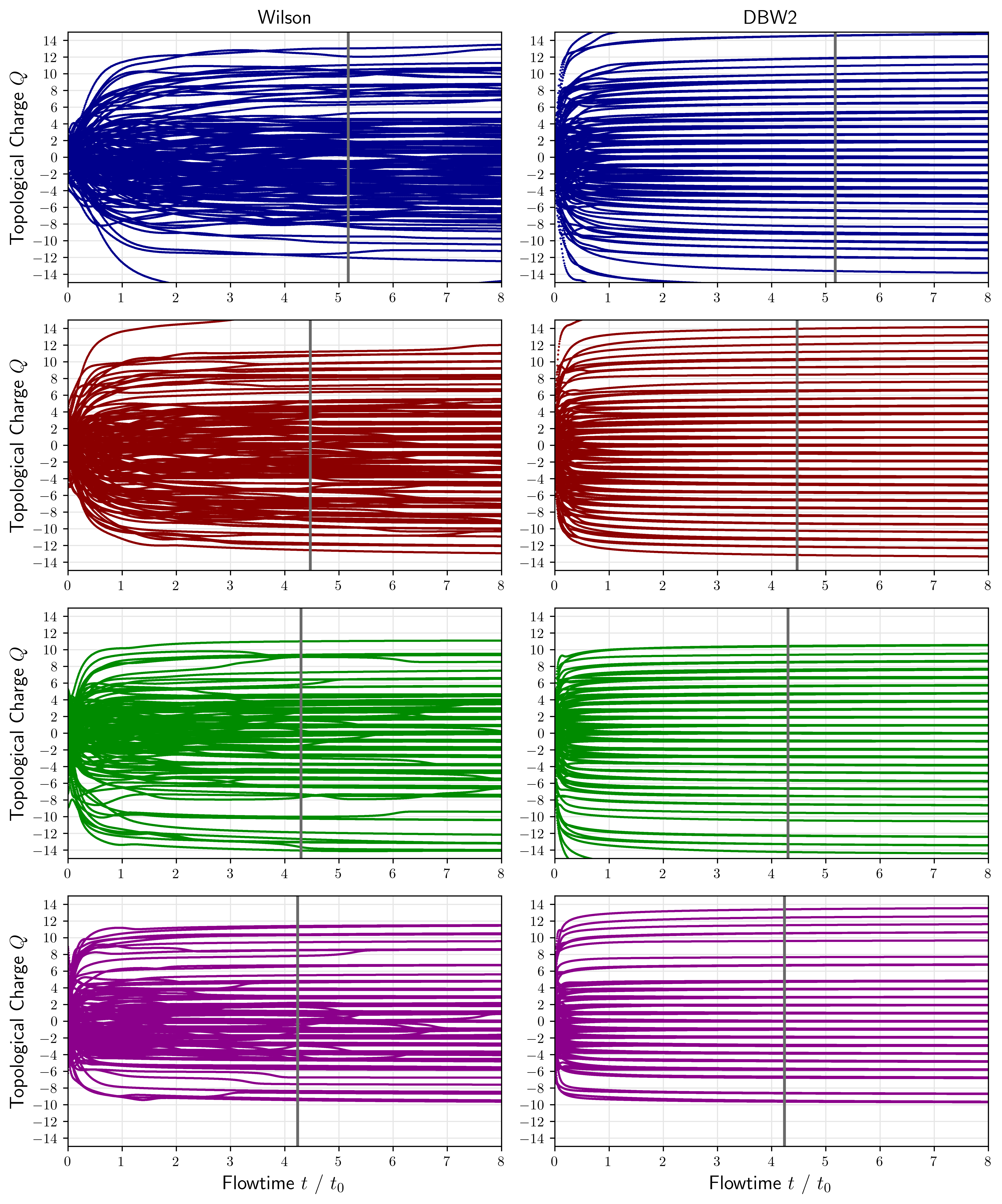}
\caption{Topological charge $Q$ as a function of flow time using the Wilson (left) and DBW2 (right) flows. Four ensembles are shown from top to bottom: $\beta = 7.1$ (top, blue), $\beta = 7.2$ (red), $\beta = 7.3$ (green), and $\beta = 7.4$ (bottom, purple). Only the first 150 configurations after thermalization are shown. The vertical grey line indicates where the flow wraps around the lattice, $\sqrt{8t} \approx L/2$.}
\label{fig.flow}
\end{figure}

Figure~\ref{fig.flow} shows the topological charge $Q$ as a function of normalised flow time $t/t_0$. The left panels correspond to the Wilson flow and the right panels to the DBW2 flow. The DBW2 flow stabilises the topological charge at early flow times ($t/t_0 \gtrsim 1$), while the Wilson flow exhibits significant fluctuations over the entire period shown. This behaviour improves with finer lattice spacings, but transitions between topological sectors within one configuration remain visible for the Wilson flow. In contrast, such transitions are absent in the DBW2 flow.

\begin{figure}%[!ht]
\centering
\includegraphics[width=1.0\textwidth]{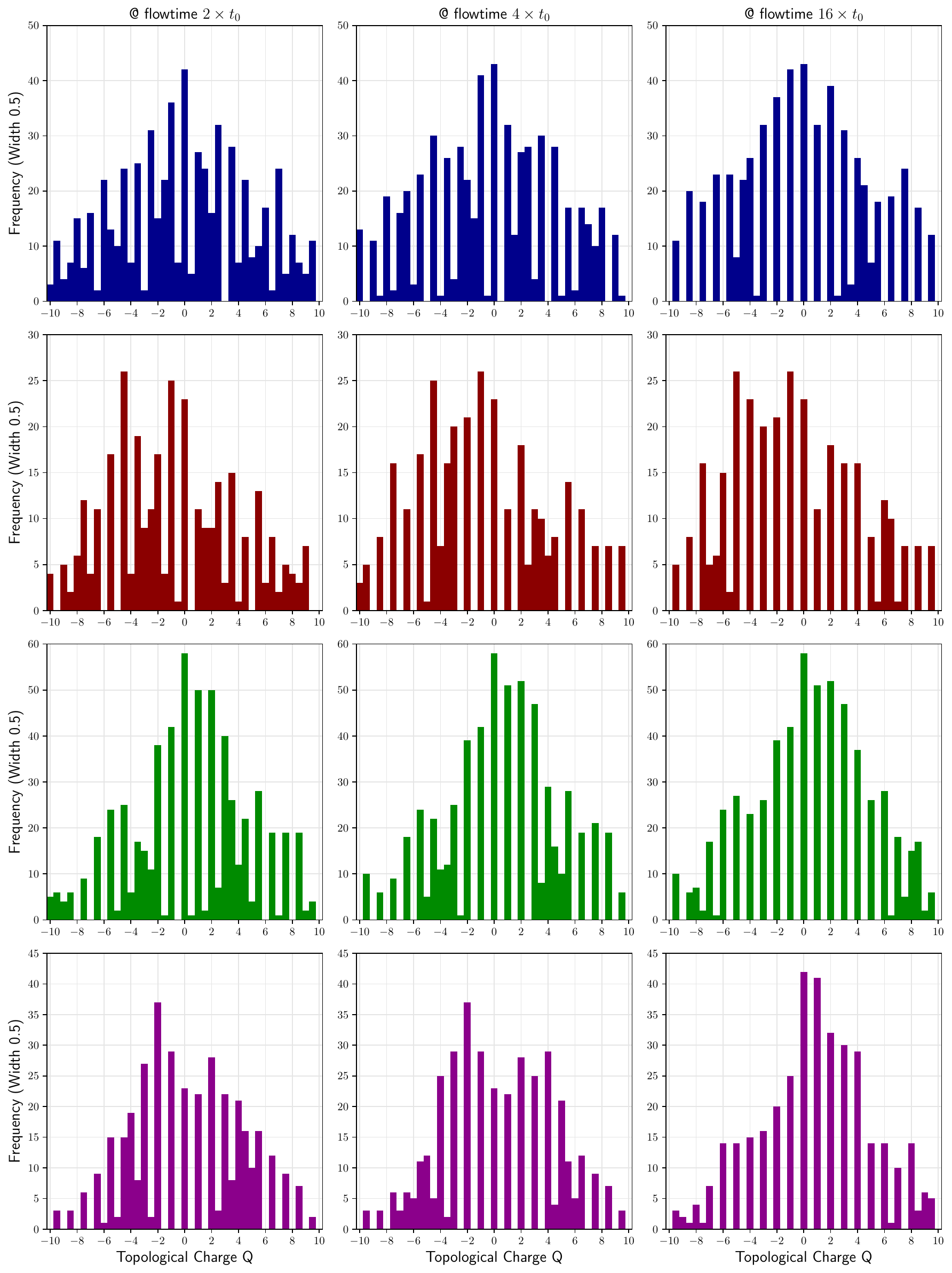}
\caption{Histograms of topological charge $Q$ for different flow times using the DBW2 flow. Each row corresponds to an ensemble, from $\beta = 7.1$ (top) to $\beta = 7.4$ (bottom). Columns show results for $2t_0$, $4t_0$, and $16t_0$.}
\label{fig.histo}
\end{figure}

\begin{figure}%[!ht]
\centering
\subfloat{
\includegraphics[width=0.499\textwidth]{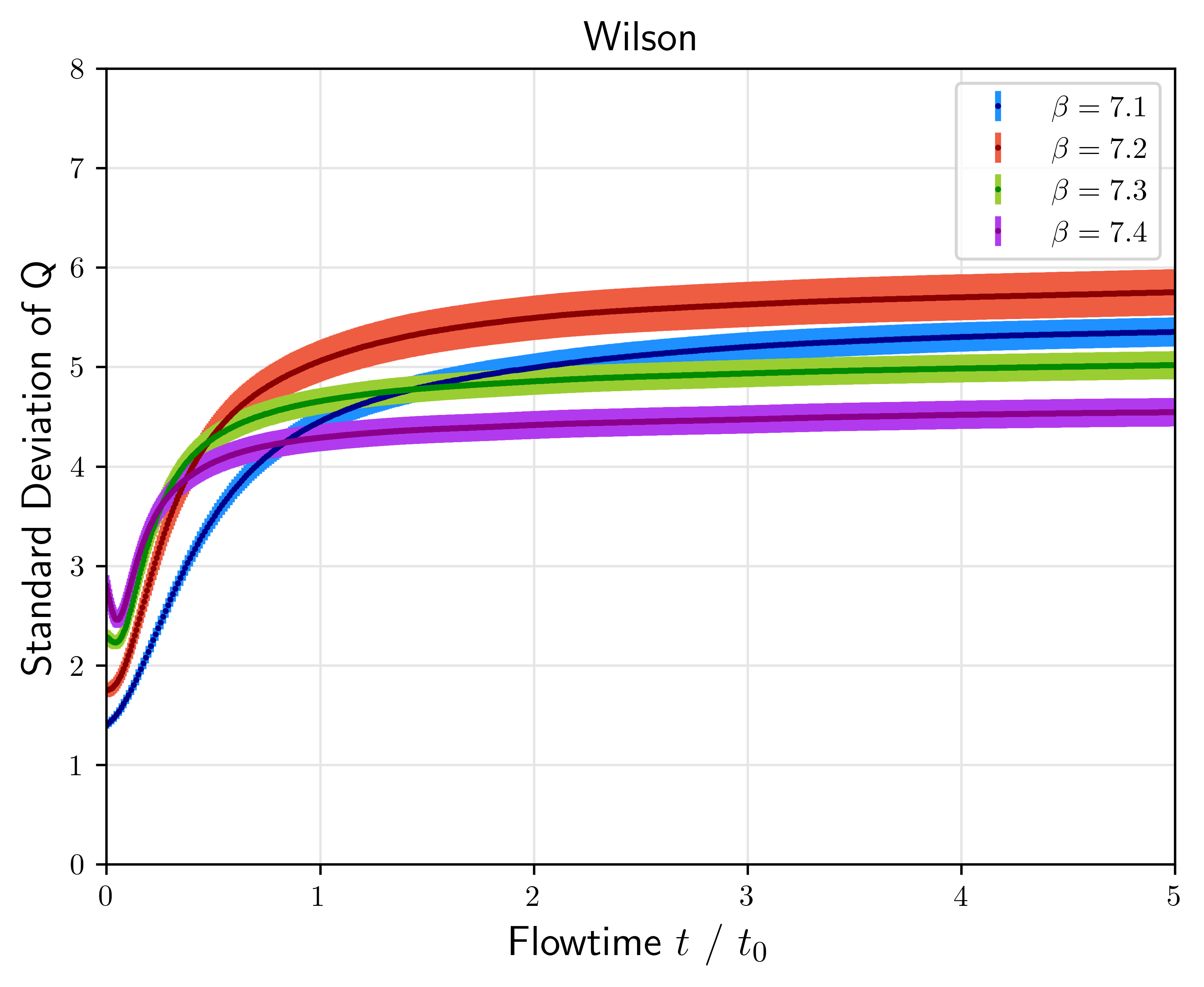}
}
\subfloat{
\includegraphics[width=0.499\textwidth]{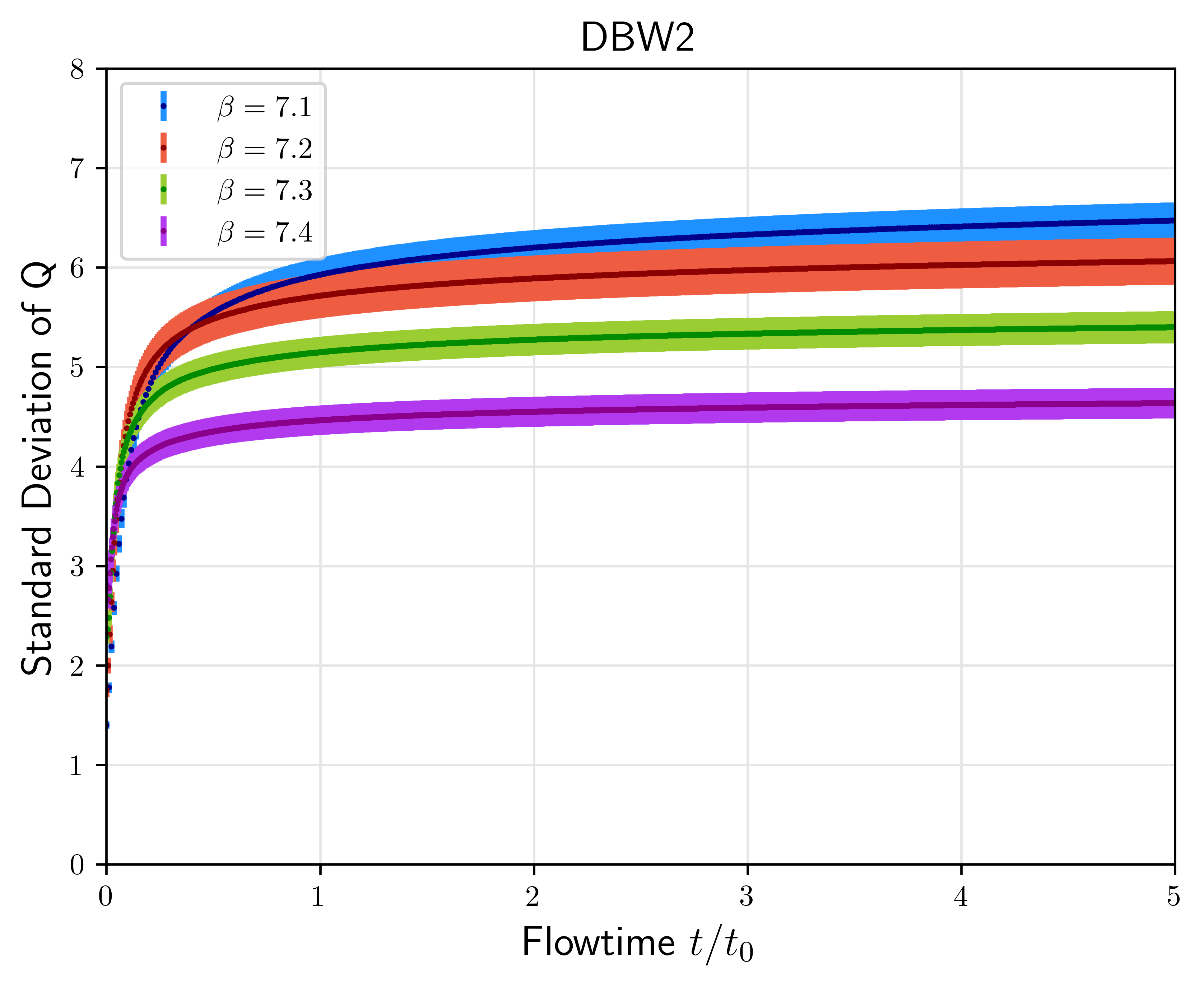}
}
\caption{Standard deviation of the topological charge as a function of the flowtime $t / t_0$ for the Wilson (left) and the DBW2 (right) flow. The four colours correspond to the different ensembles.}
\label{fig.std}
\end{figure}

Figure~\ref{fig.histo} displays histograms of the topological charge $Q$ for different flow times. Integer values of $Q$ become more distinct with either increased flow time or finer lattice spacings. 
No evidence of fractional charges is observed, except for large values of $Q$. However those may result from discretisation effects due to the non-exact integer quantisation of the topological charge on the lattice as defined here. Figure~\ref{fig.std} and table~\ref{tab.std} illustrate the standard deviation of the topological charge as a function of flow time for both the Wilson and DBW2 flows. Despite the instability of the Wilson flow in producing consistent topological charges, the standard deviation is comparable to that obtained with the DBW2 flow. The errors are purely statistical and computed using bootstrap resampling.

\section{Conclusions and Outlook}

We investigated the presence of fractional topological charges in SU(4) gauge theories with fermions in the two-index antisymmetric representation. Using the DBW2 flow at sufficiently large flow times ($t/t_0 > 2$), we found no evidence of fractional charges, even in coarse lattice setups. Future studies will extend this work to larger gauge groups ($N_C = 5, 6$), where smaller fractional charges are in general possible.

\begin{table}[!ht]
    \centering
    \begin{tabular}{cc|cccccc}
    \toprule
Flowtype & Lattice & $0 \times t_0$ & $1 \times t_0$ & $ 2 \times t_0$  & $ 4 \times t_0$ & $ 8 \times t_0$ & $16 \times t_0$\\
\bottomrule 
Wilson & $\beta = 7.1$ & 1.40(4) & 5.93(17) & 6.20(17) & 6.41(18) & 6.59(19) & 6.72(19) \\ 
 & $\beta = 7.2$ & 1.75(7) & 5.72(22) & 5.89(23) & 6.03(23) & 6.14(24) & 6.22(24) \\
 & $\beta = 7.3$ & 2.29(7) & 5.15(15) & 5.28(16) & 5.37(16) & 5.45(16) & 5.52(16) \\
 & $\beta = 7.4$ & 2.81(10) & 4.47(15) & 4.55(15) & 4.62(15) & 4.67(15) & 5.15(22) \\ 
\bottomrule 
DBW2 & $\beta = 7.1$ & 1.40(4) & 4.45(12) & 4.99(14) & 5.30(15) & 5.44(15) & 5.56(15) \\ 
 & $\beta = 7.2$ & 1.75(7) & 5.06(20) & 5.50(22) & 5.70(23) & 5.86(23) & 5.97(24) \\ 
 & $\beta = 7.3$ & 2.29(7) & 4.66(13) & 4.86(13) & 4.99(14) & 5.07(14) & 5.13(14) \\ 
 & $\beta = 7.4$ & 2.81(10) & 4.29(13) & 4.42(14) & 4.52(14) & 4.58(14) & 4.84(20) \\ 
\end{tabular}
        \caption{The standard deviation of the topological charge for Wilson and DBW2 at different flow times.}
    \label{tab.std}
\end{table}

%\vspace{-0.85cm}

\section*{Acknowledgements}

%\vspace{-0.15cm}
We thank Antonio Gonz\'alez-Arroyo for discussions and Pietro Butti for discussions and for supporting our development of the code for the DBW2 flow. As part of the EuroHPC and DeiC initiatives, the calculations were performed using the Discoverer under grant number EHPC-REG-2023R01-102 and the LUMI supercomputers under grant numbers DeiC-SDU-N5-2024055 and EHPC-EXT-2024E01-038. This project has received funding from the European Union’s Horizon 2020 research and innovation program under the Marie Sk\l{}odowska-Curie grant agreements numbers 813942 and 894103.

\bibliographystyle{apsrev4-1}
\bibliography{main.bib}

\end{document}